\documentclass[conference]{IEEEtran}
\usepackage[T1]{fontenc}

\usepackage{amsmath}
\usepackage{amssymb}
\usepackage{amsfonts}
\usepackage{graphicx}
\usepackage{epsfig}
\usepackage{subfigure}
\usepackage{psfrag}
\usepackage{cite}
\usepackage{latexsym}
\usepackage{lettrine}
\usepackage{url}
\usepackage{color}
\usepackage{multirow}

\usepackage[bottom=1.09in,top=0.73in,left=0.64in,right=0.64in]{geometry}

\usepackage{verbatim}

\usepackage{algorithm}
\usepackage{algpseudocode}
\usepackage{stfloats}
\usepackage[english]{babel}
\usepackage{amsthm}
\usepackage{titlesec}

\usepackage{bm}
\usepackage{array}
\usepackage{hyperref}					% Reference
\hypersetup{colorlinks,%				% Reference color setup	
	linkcolor=blue,%
	anchorcolor=blue,
    citecolor=blue}

\newcommand{\Expct}[1]{\mathbb{E}\left\{#1\right\}}
\newcommand{\Var}[1]{\mathbb{D}\left\{#1\right\}}
\newcommand{\Abs}[1]{\left|#1\right|}

\newcommand{\Real}[1]{\Re\left\{#1\right\}}

\newcommand{\Fourier}[1]{\mathcal{F}\left\{#1\right\}}

\newtheoremstyle{boldnote} % Name of the style
  {1pt} % Space above
  {1pt} % Space below
  {\itshape} % Body font
  {} % Indent amount
  {\bfseries} % Theorem head font
  {.} % Punctuation after theorem head
  {.5em} % Space after theorem head
  {\thmname{#1}\thmnumber{ #2}\thmnote{\bfseries{ (#3)}}} % Theorem head spec

\theoremstyle{boldnote}
\newtheorem{assumption}{Assumption}

\theoremstyle{boldnote}\newtheorem{proposition}{\textbf{Proposition}}

\begin{document}
\title{Improving the Ranging Performance of Random ISAC Signals Through Pulse Shaping Design}

\author{Zihan Liao, Fan Liu, Shuangyang Li, Yifeng Xiong, Weijie Yuan, and Marco Lops}


%Telephone: (800) 555--1212, Fax: (888) 555--1212}


\maketitle

\begin{abstract}
In this paper, we propose a novel pulse shaping design for single-carrier integrated sensing and communication (ISAC) transmission. Due to the communication information embedded in the ISAC signal, the resulting auto-correlation function (ACF) is determined by both the information-conveying random symbol sequence and the signaling pulse, where the former leads to random fluctuations in the sidelobes of the ACF, impairing the range estimation performance. To overcome this challenge, we first analyze the statistical characteristics of the random ACF under the symbol-wise pulse shaping (SWPS) regime. As a step further, we formulate an optimization problem to design ISAC pulse shaping filters, which minimizes the average integrated sidelobe level ratio (ISLR) while meeting the Nyquist criterion, subject to power and bandwidth constraints. We then show that the problem can be recast as a convex quadratic program by expressing it in the frequency domain, which can be readily solved through standard tools. Numerical results demonstrate that the proposed pulse shaping design achieves substantial ranging sidelobe reduction compared to the celebrated root-raised cosine (RRC) pulse shaping, given that the communication throughput is unchanged.
\end{abstract}


% \fi
\IEEEpeerreviewmaketitle

\vspace{-7pt}
\section{Introduction}
Integrated Sensing and Communications (ISAC) is well-recognized as a promising technology for 6G wireless networks, which is capable of supporting a variety of emerging applications including intelligent transportation, smart manufacturing, and environment monitoring \cite{liu2022integrated}. To deploy ISAC over existing networking infrastructures, a common practice is to straightforwardly adopt standard communication waveforms for sensing \cite{zhang2021overview}. Nevertheless, sensing and communication (S\&C) systems exhibit distinct signal design pipelines and thus differ in their nature of optimality. The pursuit of optimal S\&C tradeoff necessitates a comprehensive re-consideration of the respective constraints imposed on ISAC signaling and a new paradigm for ISAC signal design.
% In particular, multiple communication symbols may correspond to a single pulse in the slow-time dimension for radar transmission, where a single communication symbol could be treated as one or more snapshots in radar's fast-time dimension, depending on the number of sampling points per symbol. 

Single-carrier communication systems employ symbol-wise pulse shaping (SWPS) to produce baseband signals, where the symbol sequence is convolved with a pulse-shaping filter, such that each symbol is associated with a signaling pulse adapting to various requirements of the system \cite{xia1997family}\cite{Baas2004Pulse}. These include the out-of-band (OOB) constraint and zero inter-symbol interference constraint, where the latter is also known as the Nyquist criterion \cite{Vahlin1996Optimal}. Despite their effect of ensuring the communication performance, these constraints are not necessarily beneficial for the sensing functionality. On the other hand, conventional sensing signals are usually meticulously designed deterministic signals with favorable sensing properties, (e.g. linear frequency modulation signals), which do not carry information \cite{levanon2004radar}. This dilemma motivates us to refine the SWPS scheme by incorporating sensing-oriented design principles.

The sensing performance degradation of SWPS signals may be interpreted in light of the deterministic-random tradeoff (DRT) and the subspace tradeoff (ST) \cite{xiong2024torch}. From the perspective of DRT, the inherent randomness of communication data introduces variability in the signal, leading to changes in sensing performance metrics such as the auto-correlation function (ACF) (specifically, the zero-Doppler slice of the ambiguity function) \cite{xiao2022waveform}. This variability adversely affects the ranging accuracy, the probability of detecting multiple targets, and the false alarms rate \cite{xiao2022waveform}. From the perspective of ST, conventional SWPS designs work on communication-favorable signal subspaces, overlooking the sensing-related properties. Specifically, conventional SWPS requires specific points of the ACF to be zero (corresponding to the Nyquist criterion) \cite{Baas2004Pulse}, as will be demonstrated in later sections. However, for sensing tasks, it is beneficial to have the entire region, except for the mainlobe, maintain a low ACF value to ensure ranging and detection performance \cite{vizitiu2014some}.

The aforementioned challenges complicate the analysis and optimization under conventional sensing frameworks. Against this background, in this paper, we examine the stochastic properties of the SWPS signal ACF and introduce a novel SWPS design for single-carrier ISAC that enhances the ranging performance. We formulate an optimization problem for the ISAC SWPS filter design aimed at reducing the average integrated sidelobe level ratio (ISLR) while adhering to the Nyquist criterion, as well as power and bandwidth constraints. We demonstrate that the problem can be reformulated as a convex quadratic program with respect to the power spectrum of the pulses. Numerical results show that the proposed SWPS design significantly reduces the ranging sidelobe levels compared to the traditional root-raised cosine (RRC) pulse shaping while preserving the communication throughput.
\vspace{-5pt}

\section{Signal Model and Performance Metrics}

\subsection{Signal Model}

Within the structure of SWPS, the symbol at the $n$-th position in a communication downlink frame of length $L$, denoted as $s_n$, undergoes modulation to create a symbol pulse $s_n(t) = s_n g(t)$. Here, $g(t)$ represents the impulse response of the pulse shaping filter. Subsequently, The frame signal can be articulated as the aggregate of the time-shifted symbol pulses as below
\begin{equation}
\vspace{-5pt}
    s(t) = \sum_{n=0}^{L-1} s_n(t-nT) = \sum_{n=0}^{L-1} s_n g(t - nT),
\end{equation}
where $T$ denotes the symbol duration.

Before delving into the metrics of S\&C, we establish the following assumptions which are commonly made in practical communication systems:
\begin{assumption}[Symmetric Constellation]
\label{assump:symmetric}
In this paper, our focus is on the constellation where symbols are centrosymmetric about the origin, adhering to the principle that $\Expct{s_n^p}=0, \forall n$, where p is any positive integer. This assumption is satisfied by practical constellations such as QAM and PSK.
\end{assumption}

\begin{assumption}[Independent Symbols]
\label{assump:independent}
We assume that $s_n$ and $s_m$ are independent when $n$ does not equal $m$, leading to $\Expct{s_n s_m^*}=0$ for all instances where $n\neq m$.
\end{assumption}

\begin{assumption}[Identical Constellation]
\label{assump:identical}
We assume that all symbols within a single frame share the same constellation, whose energy is normalized to $1$, implying that
\begin{equation}
    \Expct{\Abs{s_n}^2}=1, \; \Expct{\Abs{s_n}^4}=\mu_4, \forall n,
\end{equation}
where $\mu_4\geq 1$ is a constant.
\end{assumption}

\subsection{The Auto-Correlation Function and the Sensing Metric}
Let us first give the definition of the ACF of $s(t)$ by
\begin{equation}
\begin{aligned}
\chi(\tau) &= \int s(t) s^*(t-\tau) dt.
\end{aligned}
\end{equation}

\begin{proposition}
\label{prop:ambg}
\begin{equation}
\chi(\tau) = \sum_{n=0}^{L-1} \sum_{m=0}^{L-1} s_n s_m^* G_{n,m}(\tau),
\end{equation}
where $G_{n,m}(\tau)= G(\tau+(m-n)T)$ signifies the cross-correlation function of the $n$-th and $m$-th pulses, and $G(\tau) = \int g(t) g^*(t-\tau) dt$ is the ACF of $g(t)$.
\begin{proof}
See Appendix \ref{sec:app_A}.
\end{proof}
\end{proposition}

Additionally, the ACF of $g(t)$ may be computed by calculating the inverse Fourier transform of the power spectrum of $g(t)$ as follows
\begin{equation}
\label{eq:acf_spectr}
    G(\tau) = \int |U(f)|^2 e^{j2\pi f\tau} df,
\end{equation}
where $U(f)$ is the Fourier transform of $g(t)$.

Using the concept of ACF, we may now formally define the sensing metric used in this paper, the ISLR, as follows
\begin{equation}
\begin{aligned}
L_{ISLR}\{g(t)\}=\mathbb{E}\left\{\frac{\int_{\mathcal{T}}|\chi(\tau)|^2 d \tau}{|\chi(0)|^2}\right\}.
\end{aligned}
\label{eq:islr}
\end{equation}
where $\mathcal{T}$ represents the delay region of interest.

\subsection{Communicaion Metrics: OOB and ISI}

To control the bandwidth of the transmitted signal, it is necessary to meet the specified out-of-band (OOB) constraint:
\begin{equation}
    \int_{-\infty}^{-B/2} |U(f)|^2 df + \int_{B/2}^{\infty} |U(f)|^2 df \leq \varepsilon_{OB},
\end{equation}
where $B$ is the bandwith of the signal.

Considering a Rayleigh fading channel, the representation of the matched filtered signal at the receiver side is given by
\begin{equation}
\begin{aligned}
    y(t) &= h_c s(t)*g(-t) + n(t)*g(-t) \\
    &= \sum_{n=0}^{L-1} s_n G(t - nT) + \tilde{n}(t),
\end{aligned}
\end{equation}
where $\tilde{n}(t)$ refers to the noise part of the matched filter output and $h_c$ is the channel state information. After sampling at $t=kT,k=0,1,\cdots,L-1$, we have
\begin{equation}
    y_k = h_c\sum_{n=0}^{L-1} s_n G((k - n)T) + \tilde{n}(kT),
\end{equation}
which represents the communication input-output relationship. Hence, to meet the Nyquist criterion and avoid ISI, it is necessary to have
\begin{equation}
|G(nT)|^2=0, n=\cdots,-2,-1,1,2,\cdots.
\end{equation}
According to \cite{barry2012digital}, this constraint is equivalent to 
\begin{equation}
    \sum_{m=-\infty}^{\infty} |U(f+m/T)|^2 = T,
\end{equation}
which is known as the constant folded spectrum criterion. Under the assumption $\frac{1}{2T}\leq\frac{B}{2}\leq\frac{1}{T}$, it can be recast as
\begin{equation}
\label{eq:freq_no_isi_constr}
\begin{aligned}
&|U(f)|^2 = T, 0\leq f\leq\frac{1}{T}-\frac{B}{2}, \\
&|U(f)|^2 + \Abs{U\left(\frac{1}{T}-f\right)}^2 = T, \frac{1}{T}-\frac{B}{2}\leq f\leq\frac{B}{2}.
\end{aligned}
\end{equation}

For the sake of clarity, we denote the feasible set of $|U(f)|^2$ as $\mathcal{C}$, subject to OOB constraint and Nyquist Criterion, which can be readily shown as linear in $|U(f)|^2$.

\section{Statistical Analysis of the ACF}

\subsection{Two Parts of the ACF}

To derive the statistical characteristics of the ACF, we first split $\chi(\tau)$ into two parts by
\begin{equation}
\chi(\tau) = \underbrace{\sum_{n=0}^{L-1}\left|s_n\right|^2 G_{n,n}(\tau)}_{\chi_s(\tau)}  +\underbrace{\sum_{n=0}^{L-1} \sum_{m=0, m \neq n}^{L-1} s_n s_m^* G_{n,m}(\tau)}_{\chi_c(\tau)}.
\label{eq:twopart}
\end{equation}
$\chi_s(\tau)$ describes the part contributed by the self-ambiguity of the symbols, while $\chi_c(\tau)$ describes the part contributed by the cross-ambiguity among different symbols.

\begin{proposition}
\label{prop:chi}
We have that 
\begin{align}
    &\Expct{\chi_s(\tau)}=LG(\tau), \; \Expct{\chi_c(\tau)}=0, \label{eq:chi_expct}\\
    &\Expct{\chi_s(\tau)\chi_c^*(\tau)}=0, \label{eq:zero_corr_sc}
\end{align}
which suggests that $\chi_s(\tau)$ and $\chi_c(\tau)$ are uncorrelated.
\begin{proof}
    See Appendix \ref{sec:app_B}.
\end{proof}
\end{proposition}

\begin{proposition}
\label{prop:chisc_var}
The variance of $\chi_s(\tau)$ and $\chi_c(\tau)$ can be expressed as
\begin{align}
    \Var{\chi_s(\tau)} &= L\left(\mu_4-1\right)G(\tau)^2, \label{eq:chis_var} \\
    \Var{\chi_c(\tau)} &= \sum_{|n|<L,n\neq 0} (L-|n|) G(\tau+nT)^2. \label{eq:chic_var}
\end{align}
\begin{proof}
    See Appendix \ref{sec:app_C}.
\end{proof}
\end{proposition}

For notational convenience, we denote 
\begin{equation}
    \tilde{\alpha}_n =
    \begin{cases}
        L\left(\mu_4-1\right), & n=0, \\
        L-|n|, & \text{otherwise}.
    \end{cases}
\end{equation}

\subsection{Expectation and Variance of the ACF}

According to \eqref{eq:twopart} and \eqref{eq:chi_expct}, the expectation of the $\chi(\tau)$ can be written as 
\begin{equation}
\begin{aligned}
\mathbb{E}\{\chi(\tau)\} &= \mathbb{E}\{\chi_s(\tau)\} = LG(\tau).
\end{aligned}
\end{equation}

The variance of the ACF can be expressed as 
\begin{equation}
\begin{aligned}
    &\Var{\chi(\tau)}=\Expct{\Abs{\chi(\tau)}^2}-\Abs{\Expct{\chi(\tau)}}^2 \\
    &=\Var{\chi_s(\tau)} + \Var{\chi_c(\tau)} \\
    &\quad + \underbrace{2\Real{\Expct{\chi_s(\tau)\chi_c^*(\tau)}-\Expct{\chi_s(\tau)}\Expct{\chi_c^*(\tau)}}}_{\text{cross term}}.
\end{aligned}
\end{equation}
From \eqref{eq:chi_expct} and \eqref{eq:zero_corr_sc} we have that the cross term is equal to $0$, which gives us
\begin{equation}
\label{eq:var_addition}
    \Var{\chi(\tau)} = \Var{\chi_s(\tau)} + \Var{\chi_c(\tau)} = \sum_{|n|<L} \tilde{\alpha}_n G(\tau+nT)^2.
\end{equation}

With the expectation and the variance of $\chi(\tau)$ at hand, it is convenient to compute the expectation of the squared ACF (SACF) by
\begin{equation}
\begin{aligned}
\mathbb{E}\left\{|\chi(\tau)|^2\right\} &= |\mathbb{D}\{\chi(\tau)\}|^2+|\mathbb{E}\{\chi(\tau)\}|^2 \\
& = \sum_{|n|<L} \alpha_n G(\tau+nT)^2,
\end{aligned}
\end{equation}
where $\alpha_n = \tilde{\alpha}_n + L^2\delta[n]$ and $\delta[n]$ is the discrete impulse function.

\section{ISAC Pulse Design via ISLR Minimization}
Let us assume that the term $|\chi(0)|^2$ is nearly constant, given that it represents the square of the energy of the transmitted waveform. In fact, this assumption holds accurate if $L$ is sufficiently large. By noting $G(0) = 1$ (normalized energy), and $G(nT)=0$ for non-zero integer $n$ (Nyquist criterion), we may approximate $|\chi(0)|^2$ as  $|\chi(0)|^2\approx L^2+L(\mu_4-1)=\alpha_0$.
Consequently, $L_{ISLR}\{g(t)\}$ can be roughly estimated as
\begin{equation}
\vspace{-5pt}
\begin{aligned}
L_{ISLR}\{g(t)\}&\approx\frac{1}{\alpha_0}\mathbb{E}\left\{\int_{\mathcal{T}}|\chi(\tau)|^2 d\tau\right\} \\
&= \sum_{|n|<L} \frac{\alpha_n}{\alpha_0}\int_{\mathcal{T}}|G(\tau+nT)|^2 d\tau,
\end{aligned}
\end{equation}
The ISAC pulse shaping function should thus be designed to minimize the ISLR under communication constraints, yielding
\begin{equation}
    \min_{|U(f)|^2} \; L_{ISLR}\left\{\Fourier{|U(f)|^2}\right\} \;\; \text{s.t.}\; |U(f)|^2\in\mathcal{C},
    \label{eq:design_problem}
\end{equation}
where $\Fourier{\cdot}$ denotes the Fourier transform.

Let $g_k=g(kT_s)$ be the sampled sequence of $g(t)$ of length $L_g$ with $T_s$ being the sampling period, which can be stacked into a vector $\mathbf{g}\in\mathbb{R}^{L_g}$.
By representing the non-zero part (band-limited part) of $|U(f)|^2$ by a vector $\boldsymbol{\omega}\in\mathbb{R}^{N_B+1}$, the whole sampled power density spectrum may be constructed by
\begin{equation}
    \tilde{\boldsymbol{\omega}}=
    \begin{bmatrix}
        \boldsymbol{\omega} \\
        \mathbf{0}_{L_g-2N_B-1,1} \\
        \text{flip}(\boldsymbol{\omega}[2:N_B+1])
    \end{bmatrix}=
    \mathbf{B}\boldsymbol{\omega}.
\end{equation}
where $\mathbf{B}$ is the matrix characterizing the linear transform $\boldsymbol{\omega}\rightarrow\tilde{\boldsymbol{\omega}}$.
Therefore, the sampled ACF sequence of $g(t)$ can be equivalently expressed by the IFFT of the sampled power spectrum in the form of
\begin{equation}
\boldsymbol{\psi}=\mathbf{F}^H\tilde{\boldsymbol{\omega}}=\mathbf{F}^H\mathbf{B}\boldsymbol{\omega},
\end{equation}
where $\boldsymbol{\psi}=[G_0,G_1,\cdots,G_{L_g-1}]^{\top}$ signifies the discretized ACF of $g(t)$ and is circularly symmetric, and $\mathbf{F}$ denotes the FFT matrix. Using these notations, we may simplify the formulation of the signal design problem \eqref{eq:design_problem} as follows.

\subsection{Sensing-Oriented Objective Function}
Let $\chi_u=\chi(uT_s)$. Then the discrete ISLR $L_{ISLR}\{\boldsymbol{\omega}\}$ can be written in a convex quadratic form of $\boldsymbol{\omega}$ as
\begin{equation}
\begin{aligned}
L_{ISLR}\{\boldsymbol{\omega}\}&\approx\mathbb{E}\left\{\sum_{u\in\Theta}|\chi_u|^2\right\} = \sum_{|n|<L} \frac{\alpha_n}{\alpha_0}\sum_{u\in\Theta}G_{u+nN_T}^2, \\
&= \sum_{|n|<L}\frac{\alpha_n}{\alpha_0}\sum_{u\in\Theta}\boldsymbol{\omega}^{\top}\mathbf{T}_{u+nN_T}\boldsymbol{\omega} = \boldsymbol{\omega}^{\top}\mathbf{Q}\boldsymbol{\omega},
\end{aligned}
\end{equation}
where 
\begin{equation}
\mathbf{T}_u = 
\left\{
\begin{aligned}
    &\mathbf{B}^H\mathbf{F}\mathbf{e}_u\mathbf{e}_u^H\mathbf{F}^H\mathbf{B}, \; u \leq \left\lceil L_g/2\right\rceil, \\
    &\mathbf{0}, \; u > \left\lceil L_g/2\right\rceil,
\end{aligned}
\right.
\end{equation}
and $\mathbf{e}_u$ is a selection vector with its $u$-th element being set to one while all other elements are zero. Finally, $\mathbf{Q}$ can be expressed as
\begin{equation}
\begin{aligned}
    \mathbf{Q} &= \sum_{|n|<L}\frac{\alpha_n}{\alpha_0}\sum_{u\in\Theta}\mathbf{T}_{u+nN_T},
\end{aligned}
\end{equation}
which is a semidefinite matrix.

\subsection{Communication-Oriented Constraints}
Suppose that the roll-off factor of the pulse shaping filter is $\beta$, implying $BT=(1+\beta)$. We may then denote the number of samples per symbol duration as $N_T=(1+\beta)L_g/(2N_B)$.
Then the Nyquist condition given in (\ref{eq:freq_no_isi_constr}) can be written as
\begin{equation}\label{linear_constraints}
\vspace{-5pt}
\begin{aligned}
&\omega_n = N_T, 0 \leq n \leq \left[\frac{1-\beta}{1+\beta}N_B\right], \\
&\omega_n + \omega_{\left[\frac{2N_B}{1+\beta}\right]-n} = N_T, \left[\frac{1-\beta}{1+\beta}N_B\right]+1 \leq n \leq N_B.
\end{aligned}
\end{equation}
Note that \eqref{linear_constraints} already incoporates the energy constraint in an implicit manner. The summation of all $\omega_n$ is now a fixed constant. Moreover, by noting the fact that these constraints are linear, they can be written in the form of
\begin{equation}
    \mathbf{A}\boldsymbol{\omega}=N_T\mathbf{1}.
\end{equation}
\subsection{Discretized Problem Formulation}
We may now express the ISLR optimization problem \eqref{eq:design_problem} under communication constraints as
\begin{equation}
\underset{\boldsymbol{\omega}}{\min}\; \boldsymbol{\omega}^{\top}\mathbf{Q}\boldsymbol{\omega}\;\;\text{s.t.} \; \mathbf{A}\boldsymbol{\omega}=N_T\mathbf{1},\boldsymbol{\omega}\geq \mathbf{0},
\end{equation}
which is a linearly constrained convex QP and may be very efficiently solved by off-the-shelf numerical tools.

\section{Numerical Results}

In this section, numerical results are presented to confirm the effectiveness of the proposed pulse shaping design. We assume a single-carrier ISAC transceiver with a bandwidth of $B=20\ \text{MHz}$. The downlink frame length $L$ is set as $256$. The sampling frequency (for pulse design) is established at $32B$, implying that we sample $32$ points per symbol. The chosen constellation format is 16-QAM, with a $\mu_4=1.32$. While other constellation types are feasible, the small value of $L(\mu_4-1)$ compared to $L^2$ in $\alpha_0$ suggests that the impact of different constellations is negligible.

\begin{figure}[t]
\vspace{-10pt}
    \centering
    \includegraphics[width=0.75\linewidth]{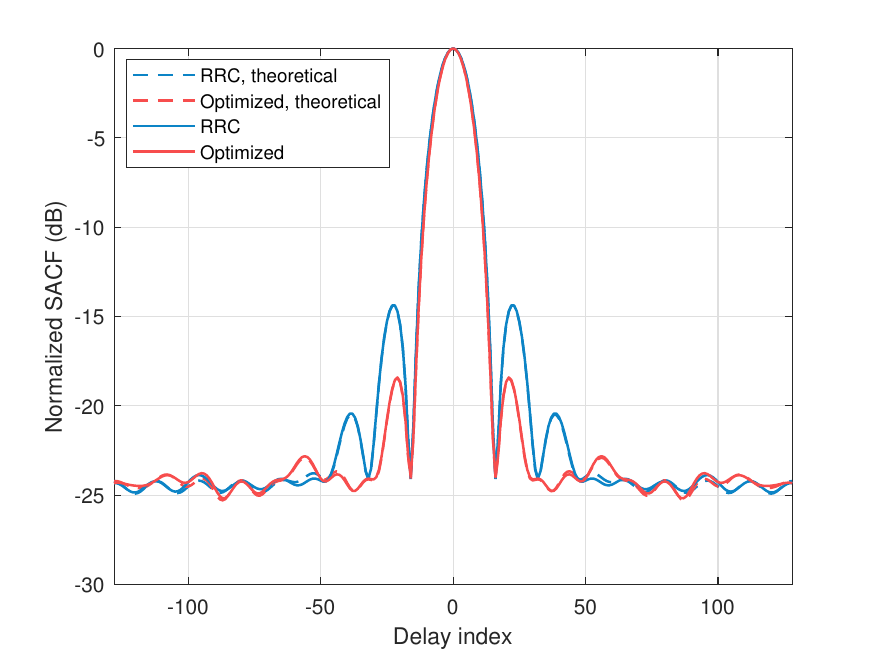}\vspace{-10pt}
    \caption{Normalized SACFs of the RRC and the optimized pulse calculated through the theoretical analysis or by averaging he SACFs of $1000$ frame signal realizations. The roll-off factor $\beta=0.3$}
    \label{fig:1}
\vspace{-10pt}
\end{figure}

Figure \ref{fig:1} displays the normalized SACFs for both the RRC and the optimized pulse for $\beta=0.3$. The SACFs presented are calculated using theoretical values derived in this paper, as well as the numerical average of $1000$ randomly realized symbol sequences. The comparison shows that the theoretical SACFs closely match their numerical counterparts. Notably, the SACF of ISAC signals with optimized pulse shaping demonstrates an approximately $4\ \text{dB}$ reduction in the first and second sidelobes compared to those shaped by the RRC.

\begin{figure}[t]
    \centering
    \includegraphics[width=0.75\linewidth]{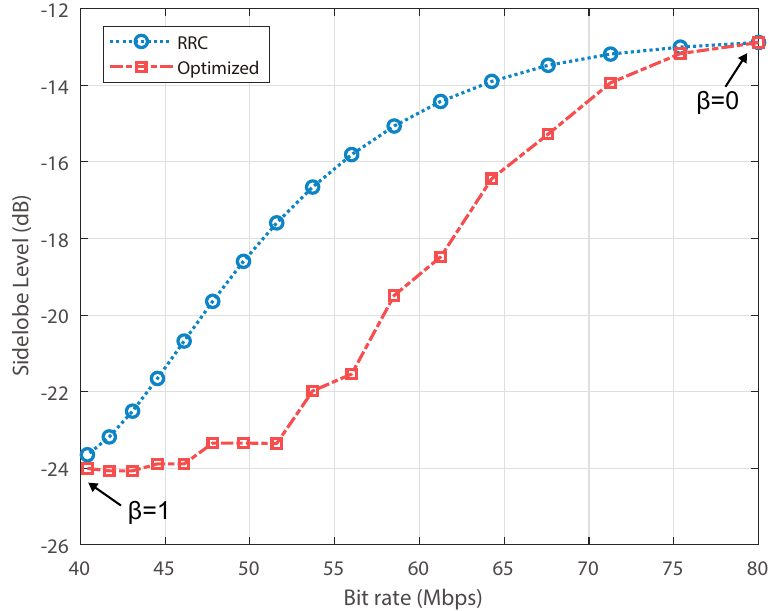}
    \vspace{-10pt}
    \caption{Second sidelobe level versus bit rate. The bit rate adjustment is achieved by altering $\beta$ from $0$ (corresponds to a bit rate of $80$ Mbps) to $1$ (corresponds to a bit rate of $40$ Mbps).}
    \label{fig:2}
\end{figure}

Figure \ref{fig:2} illustrates the tradeoff between S\&C performance by depicting how the second sidelobe level vary with the bit rate through changing $\beta$. Considering the use of 16-QAM, the bit rate may be calculated as $4/T=4B/(\beta+1)\ \text{bps}$. An increase in $\beta$ results in a decrease in bit rate and an increase in the ACF sidelobe level as shown in Fig. \ref{fig:2}. At $\beta=0$, the maximum bit rate is achieved with the sinc pulse, which is the only pulse meeting the Nyquist criterion, leading to no improvement in the S\&C tradeoff even when shaped by the optimized pulse. At $\beta=1$, the sidelobes of the ISAC signal ACF shaped by the RRC pulse are essentially flat, offering no opportunity for enhancing the S\&C tradeoff. By scaling $\beta$ from 0 to 1, the optimized design achieves up to 6 dB sidelobe reduction against the conventional RRC pulse shaping.

\begin{figure}[t]
\vspace{-5pt}
    \centering
    \includegraphics[width=0.75\linewidth]{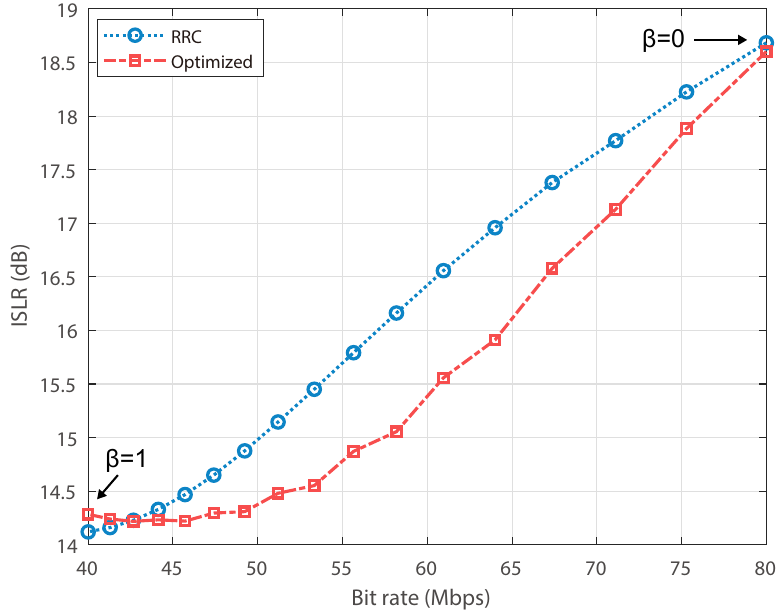}
    \vspace{-10pt}
    \caption{ISLR versus bit rate. The region of interest for the delay index of ISLR is set as $[N_T, 8N_T]$.}
    \label{fig:3}
\vspace{-10pt}
\end{figure}

Figure \ref{fig:3} demonstrates the tradeoff between ISLR and bit rate by adjusting $\beta$. As $\beta$ increases, so does the ISLR, as depicted in Fig. \ref{fig:3}. The patterns observed in this figure mirror those in Figure \ref{fig:3} and can be explained by similar reasons. At $\beta=1$ (where the bit rate is minimal), the ISLR for the RRC pulse-shaped signal is marginally lower than that for the optimized pulse, which is due to the fact that the RRC pulse does not fully satisfy the constraints owing to discretization.

\section{Conclusion}

% Utilizing a SWPS method in communications results in random variations in the ACF sidelobes, which deteriorates the accuracy of range estimation. In this paper, we first establish a model for the radar ACF within the SWPS framework, followed by an examination of the ACF's statistical behavior and its relationship with the symbol sequence and shaping pulse. Ultimately, we develop an optimization approach to design an ISAC pulse shaping filter that operates under random signaling conditions, aiming to reduce the ISLR while adhering to the constraints of communication ISI and bandwidth.

In conclusion, this paper has explored the SWPS design within the context of single-carrier ISAC systems. Identifying the significant challenges that arise from the inherent randomness of communication data, which impacts key sensing metrics, we analyze the stochastic properties of the SWPS frame signal's ACF. Subsequently, we proposed a novel SWPS design under random symbol transmission.
We formulated the SWPS filter design as a convex quadratic program that minimizes the average ISLR while satisfying the Nyquist criterion, and adhering to power and bandwidth constraints.
The numerical results confirm that the proposed design not only achieves lower ranging sidelobe levels but also maintains competitive communication throughput compared to conventional RRC pulse shaping.

\newcounter{appendixcounter}
\renewcommand\appendix{\par
    \setcounter{appendixcounter}{1} % Reset section counter
    \renewcommand{\thesection}{\Alph{appendixcounter}} % Use letters for section numbers
    \titleformat{\section} % Customizing the section title
        {\centering\sc} % Format
        {Appendix \Alph{appendixcounter}:} % Label
        {0.5em} % Space between label and title text
        {\addtocounter{appendixcounter}{1}} % Before-code
}

\appendix

\begin{figure*}[ht]
\centering
\begin{equation}
\begin{aligned}
\Expct{\chi_s(\tau)\chi_c^*(\tau)} = \sum_{n'=0}^{L-1}\sum_{n=0}^{L-1}\sum_{m=0, m \neq n}^{L-1} \Expct{\Abs{s_{n'}}^2 s_n^* s_m} G_{n',n'}(\tau) G_{n,m}(\tau) = 0 \\
\end{aligned}
\label{eq:corr_sc_expct}
\end{equation}
\vspace{-5pt}
\begin{equation}
\begin{aligned}
    \Var{\chi_c(\tau)} &= \Expct{\Abs{\chi_c(\tau)}^2} = \sum_{n=0}^{L-1}\sum_{m=0, m \neq n}^{L-1}\sum_{n'=0}^{L-1}\sum_{m'=0, m' \neq n'}^{L-1} \Expct{s_n s_m^* s_{n'}^* s_{m'}} G_{n,m}(\tau)G_{n',m'}(\tau)
    = \sum_{n=0}^{L-1}\sum_{m=0, m \neq n}^{L-1} G_{n,m}(\tau)^2
\end{aligned}
\label{eq:corr_chic_app}
\end{equation}
\hrule
\vspace{-10pt}
\end{figure*}

\section{Proof of Proposition \ref{prop:ambg}}
\label{sec:app_A}

Replacing $s(t)$ with $\sum_{n=0}^{L-1} s_n g(t-n T)$, we have
\begin{equation}
\begin{aligned}
&\chi(\tau) =\int\left(\sum_{n=0}^{L-1} s_n g(t-n T)\right)\left(\sum_{m=0}^{L-1} s_m^* g(t-m T-\tau)\right) dt \\
&=\sum_{n=0}^{L-1} \sum_{m=0}^{L-1} s_n s_m^* \int g(t-n T) g(t-m T-\tau) dt \\
&=\sum_{n=0}^{L-1} \sum_{m=0}^{L-1} s_n s_m^* \underbrace{\int g(t) g(t-(\tau+(m-n) T)) dt}_{G(\tau+(m-n)T)} \\
\end{aligned}
\end{equation}
\vspace{-20pt}

\section{Proof of Proposition \ref{prop:chi}}
\label{sec:app_B}

Recall that in Assumption \ref{assump:independent} we have $\Expct{s_n s_m^*}=0,\forall n\neq m$ and that in Assumption \ref{assump:identical} we have $\Expct{\Abs{s_n}^2}=1,\forall n$, thus
\begin{equation}
\begin{aligned}
    &\Expct{\chi_s(\tau)} = \sum_{n=0}^{L-1} \Expct{\Abs{s_n}^2} G_{n,n}(\tau) = LG(\tau), \\
    &\Expct{\chi_c(\tau)} = \sum_{n=0}^{L-1} \sum_{m=0, m \neq n}^{L-1} \Expct{s_n s_m^*}G_{n,m}(\tau) = 0.
\end{aligned}
\end{equation}

Recall that in Assumption \ref{assump:symmetric} we have $\Expct{s_n}=0,\forall n$. Thus for any tuple $\left(n',n,m\right)$ where $n\neq m$, we have that
\begin{equation}
\begin{aligned}
&\Expct{\Abs{s_{n'}}^2 s_n^* s_m}=\\
&\left\{
\begin{aligned}
    &\Expct{\Abs{s_{n}}^2 s_n^*}\Expct{s_m}=0, &&n'=n\neq m, \\
    &\Expct{\Abs{s_{m}}^2 s_m}\Expct{s_n}^*=0, &&n'=m\neq n, \\
    &\Expct{\Abs{s_{n'}}^2}\Expct{s_n}^*\Expct{s_m}=0, &&n'\neq m, n'\neq n.
\end{aligned}
\right.
\end{aligned}
\end{equation}
Then according to \eqref{eq:corr_sc_expct} we have $\Expct{\chi_s(\tau)\chi_c^*(\tau)}=0$.

\section{Proof of Proposition \ref{prop:chisc_var}}
\label{sec:app_C}

Recall that we assume symbols to be independent, which means that
\begin{equation}
\vspace{-5pt}
    \Expct{\Abs{s_n}^2\Abs{s_m}^2} = \Expct{\Abs{s_n}^2}\Expct{\Abs{s_m}^2} = 1, \forall n\neq m.
\end{equation}
In Assumption \ref{assump:identical} we have $\Expct{\Abs{s_n}^4}=\mu_4,\forall n$, thus
\begin{equation}
\begin{aligned}
    &\Var{\chi_s(\tau)} = \Expct{\Abs{\chi_s(\tau)}^2} - \Abs{\Expct{\chi_s(\tau)}}^2 \\
    &= G(\tau)^2\sum_{n=0}^{L-1}\sum_{m=0}^{L-1} \Expct{\Abs{s_n}^2\Abs{s_m}^2} - L^2G(\tau)^2 \\
    &= \Big(\underbrace{L\mu_4}_{n=m} + \underbrace{(L^2-L)}_{n\neq m} - L^2\Big)G(\tau)^2 = L(\mu_4-1)G(\tau)^2.
\end{aligned}
\label{eq:corr_chis_app}
\end{equation}

Recall that in Assumption \ref{assump:symmetric} we have $\Expct{s_n^2}=0,\forall n$. Thus for any tuple $\left(n,m,n',m'\right)$ where $n\neq m, n'\neq m$,
\begin{equation}
\begin{aligned}
&\Expct{s_n^* s_m s_{n'}^* s_{m'}} = \\
&\left\{
\begin{aligned}
    &\Expct{\Abs{s_{n}}^2}\Expct{\Abs{s_{m}}^2}=1, &&n=n',m=m',n\neq m, \\
    &\Expct{s_{n}^2}\Expct{s_{m}^2}=0, &&n=m',m=n',n\neq m. \\
\end{aligned}
\right.
\end{aligned}
\end{equation}
Then according to \eqref{eq:corr_chic_app} we have
\begin{equation}
\begin{aligned}
    \Var{\chi_c(\tau)} &= \sum_{n=0}^{L-1}\sum_{m=0, m \neq n}^{L-1} G(\tau+(m-n)T,\nu)^2 \\
    &= \sum_{|n|<L,n\neq 0} (L-|n|) G(\tau+nT)^2.
\end{aligned}
\end{equation}

\bibliographystyle{IEEEtran}
\bibliography{bare_conf}

\end{document}